# Optical polarization and intervalley scattering in single layers of $MoS_2$ and $MoSe_2$

G. Kioseoglou[a,b], A.T. Hanbicki[c], M. Currie[d], A.L. Friedman[c], and B.T. Jonker[c]

[a] *Department of Materials Science and Technology, University of Crete, Heraklion Crete, 71003, Greece*

[b] *Institute of Electronic Structure and Laser (IESL), Foundation for Research and Technology Hellas (FORTH), Heraklion Crete, 71110, Greece*

[c] *Materials Science & Technology Division, Naval Research Laboratory*, Washington, DC 20375

[d] *Optical Sciences Division, Naval Research Laboratory*, Washington, DC 20375

Single layers of $MoS_2$ and $MoSe_2$ were optically pumped with circularly polarized light and an appreciable polarization was initialized as the pump energy was varied. The circular polarization of the emitted photoluminescence was monitored as function of the difference between the excitation energy and the A-exciton emission at the K-point of the Brillouin zone. Our results show a threshold of twice the LA phonon energy, specific to the material, above which phonon-assisted intervalley scattering causes depolarization. In both materials this lead to almost complete depolarization within ~100 meV above the threshold energy. We identify the extra kinetic energy of the exciton (independent of whether it is neutral or charged) as the key parameter for presenting a unifying picture of the depolarization process.

Materials often exhibit fundamentally new phenomena in reduced dimensions, and the new properties that emerge can lead to novel applications. Like graphite, the molybdenum-based transition metal dichalcogenides, $MoX_2$ (X = S, Se) are a class of materials that readily lend themselves to dimensional manipulation. These layered structures have strong intralayer bonding and weak interlayer van der Waals coupling, enabling one to isolate individual layers. In bulk form, $MoX_2$ are indirect gap semiconductors. However, because of their reduced dimensionality, single layers are direct-gap with a range of bandgaps in the visible regime. This makes them ideal candidates for a host of optoelectronic applications ranging from light-emitting diodes to light harvesting to sensors[1-5]. Besides the obvious light-based applications, these materials are also good candidates for the emerging field of valleytronics[6-12]. The reduced-dimensional, hexagonal lattice leads to non-degenerate K-points in the Brillouin zone making the inequivalent K and K' valley populations potential new state-variables of these systems.

A single layer of $MoX_2$, schematically shown in Fig. 1(a), consists of a plane of transition metal atoms sandwiched by layers of chalcogen atoms. The Brillouin zone of this system is shown in Fig. 1(b). In single layer form, the direct gap occurs at the K-point, and the optical properties of these materials are governed by strong excitonic transitions, both neutral and charged. These excitons are strongly bound, with the neutral exciton having binding energies on the order of 0.5 - 0.8 eV[13-17] and the charged exciton (trion) having a binding energy of 20-30 meV[9,18]. In addition to being optically active with strong photoluminescence (PL), they also have unique optical selection rules (Fig. 1(c)). Time reversal symmetry and strong orbital-hybridization couples the valley and spin indices. This enables access to a single valley using polarized light, since the angular momentum of incident circularly-polarized light interacts with carriers to produce specific spin states[8-11, 19,20]. In other words, the symmetry properties of the two inequivalent valleys lead to a difference in the absorption of circularly polarized light, either positive ($\sigma^+$) or negative ($\sigma^-$) helicity, resulting in a strong chiral selectivity.

Polarization-resolved photoluminescence of single layer $MoS_2$ has been extensively studied, and a high initial circular polarization was reported with a circularly polarized pump[8,19]. In contrast, there are few polarization-dependent studies for $MoSe_2$ at zero magnetic field. One recent study[21] reported very low initial circular polarization in the PL. This result is surprising since one expects members of the $MoX_2$ family to behave similarly. Several other recent studies[22,23], demonstrated generation of valley polarization in $MoSe_2$ by applying external magnetic fields up to 10 Tesla perpendicular to the sample. In particular, MacNeil et al.[22] observed a maximum polarization of 14% for the charged and 9% for the neutral exciton at 4.2 K and 6.7 Tesla attributed to the magnetic field breaking the K/K' valley-degeneracy.

In this work, we measure the energy-dependent valley polarization in both $MoS_2$ and $MoSe_2$ under zero magnetic field, and model the polarization relaxation. We probe the valley population dynamics in $MoSe_2$ and $MoS_2$ by selectively populating the K and K' valleys with circularly polarized light while systematically varying the laser excitation energy. For both systems, the difference in the excitation energy and PL emission energy, $\Delta E = E_{pump} - E_{PL}$, governs the depopulation of carriers in each valley. Adding more energy above a distinct threshold characteristic of the longitudinal acoustic (LA) phonon for each material enables inter-valley scattering and produces a sharp decrease in the observed circular polarization. LA phonons in these two systems have different energies (30 meV for $MoS_2$ and 19 meV for $MoSe_2$)[24,25], and we show that the threshold for the excess energy required to initiate the depolarization process clearly reflects the material specific phonon energy. In addition, our results show that independent of how many carriers are excited, i.e. whether you create neutral or charged excitons, the scattering process is the same. We find that the key parameter for the depolarization process is the extra kinetic energy of the exciton – depolarization is due to intervalley scattering that begins to occur when the exciton energy exceeds a threshold corresponding to twice the LA phonon energy.

**Results**

Flakes of $MoS_2$ and $MoSe_2$ mechanically exfoliated from bulk crystals were used in this study. Monolayer $MoX_2$ regions were identified with an optical microscope (Fig. 2(a) and 2(b)) and confirmed with Raman spectroscopy at room temperature (Fig. 2(c) and 2(d)). Raman spectroscopy confirms single-layer regions through the shapes and relative positions of the out-of-plane $A_{1g}$ and in-plane $E_{2g}$ Raman active modes[26-28] (see Methods section). In Fig.2(c) the 18 cm$^{-1}$ splitting between $E^1_{2g}$ (384 cm$^{-1}$) and $A_{1g}$ (402 cm$^{-1}$) modes verifies the monolayer nature of the $MoS_2$ sample[26]. Raman spectra from $MoSe_2$ taken under the same conditions are shown in Fig. 2(d) for several layer thicknesses. The identification of the single layer[27] is based on the absence of the $B_{2g}$ mode at 353 cm$^{-1}$, which can be clearly seen in Fig. 2(d). A micro-PL setup was used to collect the PL in a backscattering geometry (see methods section). The PL spectra were analyzed as $\sigma^+$ and $\sigma^-$ using a combination of quarter-wave plate (liquid crystal retarder) and linear polarizer placed before the spectrometer entrance slit. The degree of circular polarization is defined as $P_{circ} = (I_+ - I_-) / (I_+ + I_-)$, where $I_+$ ($I_-$) is the intensity of the $\sigma^+$ ($\sigma^-$) component of the PL.

Temperature dependent PL emission from $MoS_2$ and $MoSe_2$ are shown in Fig. 3(a) and Fig. 3(b), respectively (full experimental details are described elsewhere[11]). In these spectra, the samples were excited with a laser at 2.33 eV (532 nm). The dominant emission peak is the A-exciton, a feature that originates from the lowest energy transition at the K-point of the Brillouin zone (see Fig. 1(c)). At low temperature, we measure the A-exciton at 1.89 eV for $MoS_2$, and at 1.625 eV for $MoSe_2$ reflective of the smaller bandgap of $MoSe_2$. The width of the emission from $MoS_2$ is very broad (FWHM = 0.09 eV) so it is not possible to distinguish features within the spectra and we assign this feature to the neutral exciton, $X^0$. However, the emission from $MoSe_2$ is much narrower (FWHM = 0.01 eV), and it is possible to distinguish two clear peaks separated by 0.03 eV that become apparent as the sample is heated (Fig.3 (b)). These peaks have been identified as a neutral exciton ($X^0$) and a charged exciton (T)[18]. It is not possible to determine whether the lower energy peak is a positive or negative charged exciton based solely on the emission energy, because the effective mass of electrons and holes in this material is similar.

As the temperature is increased the relative luminescence of charged *vs.* neutral exciton changes, with the neutral exciton eventually becoming the dominant feature. Note that the charged exciton is still visible at temperatures much higher than is seen in other quasi-2D systems such as GaAs QWs[29]. The temperature dependence of the exciton emission channels for both systems shows a typical semiconductor behavior (Fig. 3 (c) and 3 (d), solid symbols) and the data can be fit using a standard hyperbolic cotangent relation as defined by O'Donnell and Chen[30]. The solid lines are the fits to the data with the function $E(T) = E(0) - S\langle\hbar\omega\rangle[\coth(\langle\hbar\omega\rangle/2kT) - 1]$ where $E(0)$ is the energy position of a given feature at zero temperature, $S$ is a dimensionless coupling constant, and $\langle\hbar\omega\rangle$ is an average phonon energy. The fitting parameters obtained here are $S = 1.16$ (2.18) and $\langle\hbar\omega\rangle = 19$ meV (14.5 meV) for $MoS_2$ ($MoSe_2$).

To examine the valley spin dynamics, we measure the helicity dependent PL from these monolayers. Because of the optical selection rules (Fig. 1(c)), when pumped with light of positive ($\sigma^+$) or negative ($\sigma^-$) helicity, either the K or K' valley will be selectively populated[6]. Information on the depolarization process comes from analyzing the polarization of the subsequent PL. Figures 4(a) and 4(b) show representative PL spectra analyzed for $\sigma^+$ (solid red line) and $\sigma^-$ (dashed blue line) at T = 5 K for $MoS_2$ and $MoSe_2$, respectively, at selected $\sigma^+$ excitation energies. By using a combination of sharp, long-pass filters, stray light was suppressed at the spectrometer entrance. In addition, as the excitation energy approached the PL emission energy, the excited Raman modes were cut-off from entering the detection system. This experimental setup limited our lowest excitation energies to 1.984 eV (625 nm) for $MoS_2$ and 1.722 eV (720 nm) for $MoSe_2$. It is clear from Fig. 4 that, for both materials, the higher the excitation energy, the lower the polarization of the emission[31].

The degree of circular polarization is shown in Fig. 5 as a function of the excess energy, $\Delta E = E_{pump} - E_{PL}$, the difference between the excitation energy $E_{exc}$ and PL emission energy, $E_{PL}$ (the inset of Fig. 5 is a graphical representation). Data are plotted for $MoSe_2$ (solid red circles) and $MoS_2$ (solid blue circles) and are derived from spectra where the temperature was held constant and the laser energy was varied, or the laser energy was fixed and the emission energy was varied via a change in temperature[8,11,19]. Since this plot incorporates the energy difference between excitation and PL emission rather than the specific energy of the neutral and charged excitons, the behavior observed for both materials can be shown. Data for $MoS_2$ from the literature are also plotted in Fig. 5. The open circles and the open triangles are data from Refs 8 and 20 (respectively) taken at 5 K, and the open squares are data from Ref 19 obtained at fixed excitation energy but at different temperatures. All data follow the same depolarization trend line when plotted as a function of excess energy. Using this methodology, all of the data collapse onto a single curve for each material, independent of whether the polarization of the trion or neutral exciton is considered. The data clearly demonstrate that as the excess energy increases the emitted circular polarization decreases.

**Discussion**

To explain this behavior, we begin by noting that due to the optical selection rules[6] intra-valley scattering cannot result in a reduction in the observed polarization even if the pump energy exceeds the spin-orbit splitting (160 meV for $MoS_2$, 180 meV for $MoSe_2$)[8]. Therefore, inter-valley scattering is required to account for the reduced polarization observed, and the change in momentum necessary for such scattering implicates a phonon-mediated process[8,11]. Close to resonance the emitted circular polarization is expected to be essentially 100%, since there is not enough energy in the system to facilitate intervalley scattering. As the laser excitation energy increases or as the temperature changes for a fixed pumping energy, the available excess energy $\Delta E$ increases and phonon-assisted scattering is enabled above some material-dependent energy threshold. For $MoS_2$, the combined data set clearly interpolate to 100% polarization at $\Delta E$ = 60 meV, corresponding to twice the LA phonon energy.[11]

The $MoSe_2$ data (solid red symbols) exhibit a similar behavior – the measured polarization rapidly increases as the excess energy $\Delta E$ decreases. The narrow linewidths in $MoSe_2$ allow us to distinguish the particular emission channels $X^0$ and T. Therefore, for $MoSe_2$, we can plot the polarization of both the neutral and charged exciton. Note that due to the low intensity of the $X^0$ data, it is difficult to see the polarization on the same scale as the trion emission, therefore a typical set of $X^0$ emission spectra are shown in the inset of Fig. 4. As with MoS$_2$, all the data coalesce onto the same depolarization curve as a function of excess energy. The solid red line is a model fit described below, and intercepts 100% polarization at $\Delta E = 38$ meV, corresponding to twice the MoSe$_2$ LA phonon energy of 19 meV from the literature[25]. These data make it clear that depolarization and intervalley scattering are governed by the excess energy, $\Delta E$, imparted to the photoexcited carriers through optical pumping.

To model this behavior, we begin with a familiar rate equation model in which the emitted circular polarization can be expressed as $P_{circ} = 1/(1 + 2\frac{\tau_r}{\tau_s})$, where $\tau_r$ is the exciton lifetime and $\tau_s$ the intervalley scattering time[8,11]. The increase in the excess energy $\Delta E$, leads to an increase of the phonon population. Assuming $\tau_r$ to be linearly dependent on temperature[32] and the scattering rate $\tau_s^{-1}$ to be proportional to the phonon population $\langle n \rangle = 1/(e^{\hbar\omega_q/kT} - 1)$, we can fit the data using the relation $P = 1/(1 + C\Delta E/(e^{\hbar\omega_q/(\Delta E - \hbar\omega_q)} - 1))$, where $\hbar\omega_q$ is twice the LA phonon energy, the minimum energy necessary for the exciton (electron and hole) to scatter from one valley to the other and reduce the optical polarization. The only fitting parameters are then a scaling constant $C$ and $\hbar\omega_q$. The solid lines in Fig. 5 are fits to the data and yield values for 2LA of 54 meV for MoS$_2$ and 38 meV for the MoSe$_2$, in good agreement with the respective literature values of 60 and 38 meV[24,25].

In summary, at zero magnetic field, we initialized circular polarization in MoS$_2$ and MoSe$_2$ using energy-dependent circularly-polarized optical pumping and measured the valley polarization process as a function of the excess energy absorbed by the carriers. Independent of the emission channel or the material studied (MoS$_2$ or MoSe$_2$), all the data can be modeled by a single depolarization mechanism, intervalley scattering mediated by LA phonons. The threshold needed for depolarization is found to be twice the LA phonon energy of the corresponding material, and the exciton kinetic energy greater than this is the key parameter for the depolarization. Generating high chirality photoluminescence in Mo-based two-dimensional structures enables applications in valley-photonics.

## Methods

**Sample preparation and characterization.** All samples used in this study were mechanically exfoliated from bulk crystals. The $MoS_2$ was deposited onto a 285-nm $SiO_2$ layer on a Si substrate, while the $MoSe_2$ flakes were deposited on a 90-nm $SiO_2$ layer on a Si substrate. The typical size of monolayer regions are 5–10 μm across for $MoS_2$ and 1–2 μm for $MoSe_2$ and were identified with an optical microscope and confirmed with Raman spectroscopy at room temperature and 488nm excitation. Raman spectroscopy has been established as a reliable tool for determining the specific number of layers in transition metal dichalcogenides[26-28]. In $MoS_2$, the 18 $cm^{-1}$ energy difference between the two main vibrational modes, $E^1_{2g}$ at 384 $cm^{-1}$ and $A_{1g}$ at 402 $cm^{-1}$, is the fingerprint for the accurate determination of the single layer. In $MoSe_2$, the in-plane $E^1_{2g}$ (287 $cm^{-1}$) and the out-of-plane $A_{1g}$ (242 $cm^{-1}$) modes are much lower in energy than in $MoS_2$ because of the larger mass of the Se atom. Also, the in-plane and out-of-plane modes switch positions relative to the $MoS_2$, i.e, the out-of-plane mode is softer than the in-plane one. This can be verified by measuring thicker layers of $MoSe_2$ and observing a Davydov splitting in the $A_{1g}$ mode for multi-layer flakes. The single layer is identified by the absence of the $B_{2g}$ mode at 353 $cm^{-1}$.

**Optical measurements.** The photoluminescence data were taken in a backscattering geometry using a micro-PL setup (spatial resolution of 1 μm) with a 50x objective and incorporating a continuous-flow He-cryostat. The $MoS_2$ samples were excited with either a continuous-wave 2.33 eV (532 nm) solid-state laser or a tunable pulsed laser while the $MoSe_2$ flakes were excited by a continuous-wave Ti:Sapphire laser. The pulsed source was an optical parametric amplifier (pumped by a Ti:Sapphire laser) tunable from 1.77–2.48 eV (700-500 nm) at a 250-kHz repetition rate with a double-pass grating (500 g/mm) geometry to reduce the spectral bandwidth to < 5 meV (1 nm). The photoluminescence emission was collected, passed though a polarization analyzer, and dispersed by a single monochromator equipped with a multichannel charge coupled device (CCD) detector.

## Author Information
**Correspondence and requests for materials should addressed to**
G. K.(gnk@materials.uoc.gr) or A. H. (hanbicki@nrl.navy.mil)

## Acknowledgments
We would like to thank Jim Culbertson for assistance with Raman measurements. GK gratefully acknowledges the hospitality and support of the Naval Research Laboratory where the experiments were performed. This work was supported by core programs at NRL and the NRL Nanoscience Institute, and by Air Force Office of Scientific Research under contract number F4GGA24233G001.

## Author Contributions
G.K., A.H., and M.C. performed the experiments. G.K., A.H., and M.C. analyzed the data. A.F. fabricated the samples. All authors discussed the results and G.K., A.H., M.C. and B.J. wrote the manuscript.

**Additional Information**
Competing financial interests: The authors declare no competing financial interests.

**Figure captions**

**Figure 1** (a) Crystal structure of a single-layer transition-metal dichalcogenide. A central layer of transition metal atoms is sandwiched by layers of chalcogens. (b) Brillouin zone for a reduced dimensional hexagonal lattice. Some relevant high symmetry points are indicated. (c) Schematic of the single-particle band structure at K and K′ valleys. The arrows denote the A (solid line) and B (dashed line) excitons. Each valley can only be excited with a specific helicity, σ+ or σ–.

**Figure 2** Optical image of (a) $MoS_2$ and (b) $MoSe_2$ samples. (c) Raman spectrum of single layer $MoS_2$. The 18 $cm^{-1}$ energy separation between the in-plane and out-of-plane modes is characteristic of single layer $MoS_2$. (b) Raman spectrum of 1, 2, and 3 layers of $MoSe_2$. Spectra are offset for clarity. The absence of the $B_{2g}$ mode is the fingerprint for the single layer. All spectra were taken at 300 K with an excitation wavelength of 488 nm.

**Figure 3** Normalized PL spectra for (a) $MoS_2$ and (b) $MoSe_2$ taken with unpolarized 532 nm excitation at selected temperatures. Spectra are offset for clarity. Temperature dependent shift of the exciton emission energy in (c) $MoS_2$ and (d) $MoSe_2$.

**Figure 4** PL spectra at T = 5 K analyzed for $σ^+$ and $σ^-$ at selected photo-excitation energies for (a) $MoS_2$ and (b) $MoSe_2$. The pumping is $σ^+$ in both cases. A dramatic decrease in the emitted circular polarization is observed as the excitation energy increases. The peaks were normalized and vertically shifted for clarity. For the $MoSe_2$ spectra, the inset is an enlargement of the area indicated by a rectangular box and shows the neutral exciton emission.

**Figure 5** Degree of circular polarization of the emitted PL as function of excess energy, ΔE. Data are derived from spectra where the temperature was held constant and the laser energy was varied, or the laser energy was fixed and the emission energy was varied via a change in temperature. Solid symbols (MoSe$_2$ is red and MoS$_2$ is blue) are our data, and open symbols are data from references 8,19,20. A graphical definition of excess energy is presented in the inset.

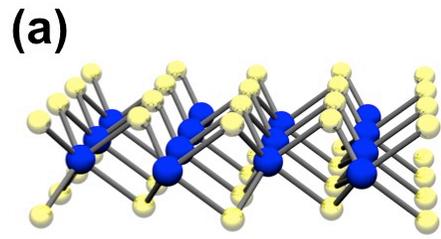
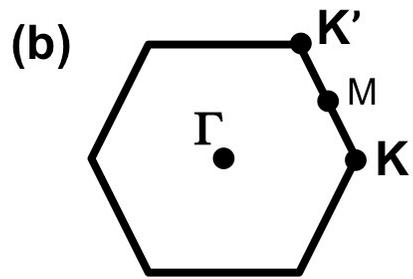
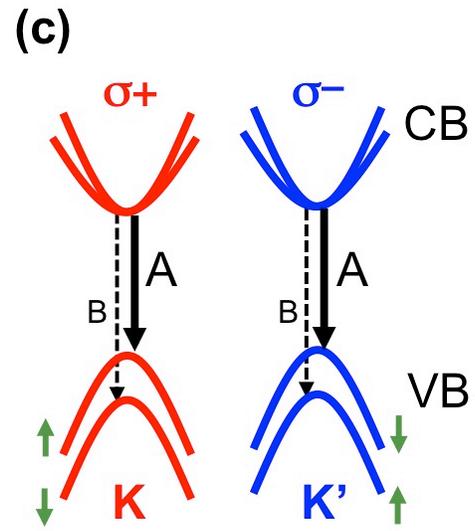

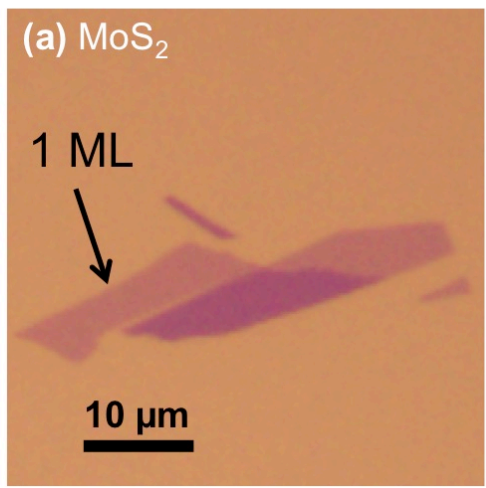
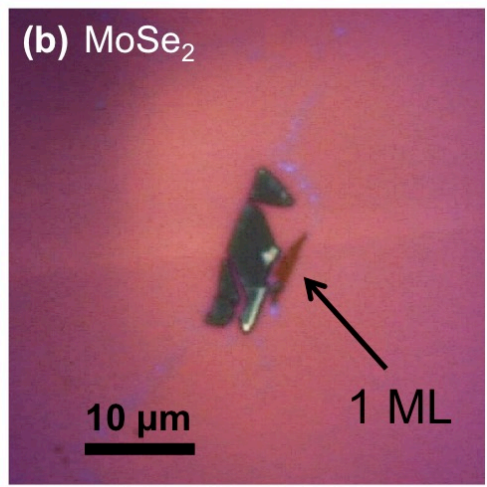
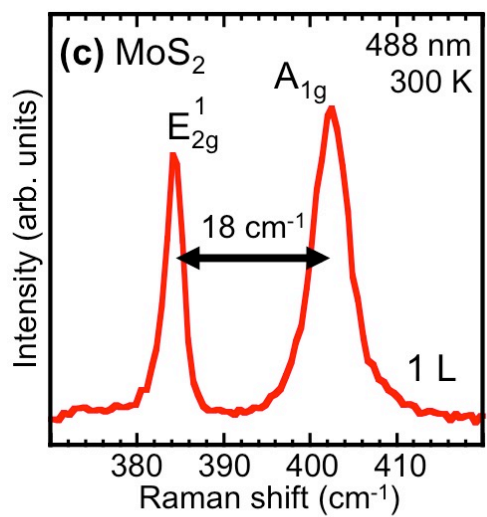
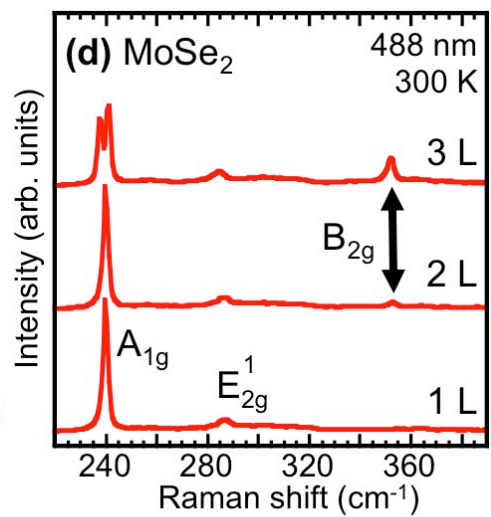

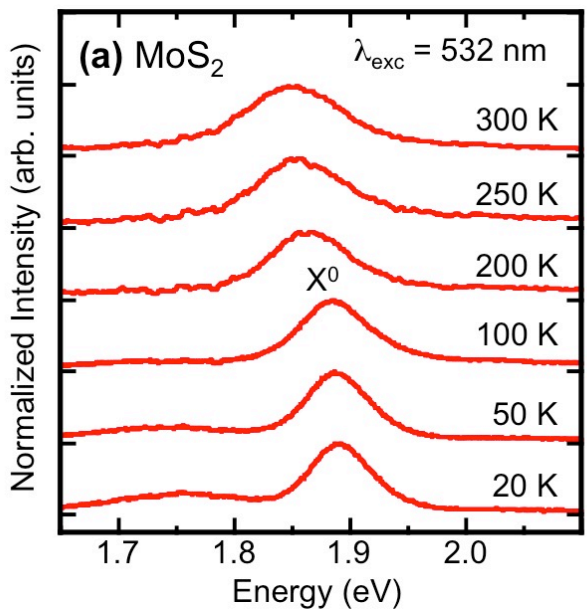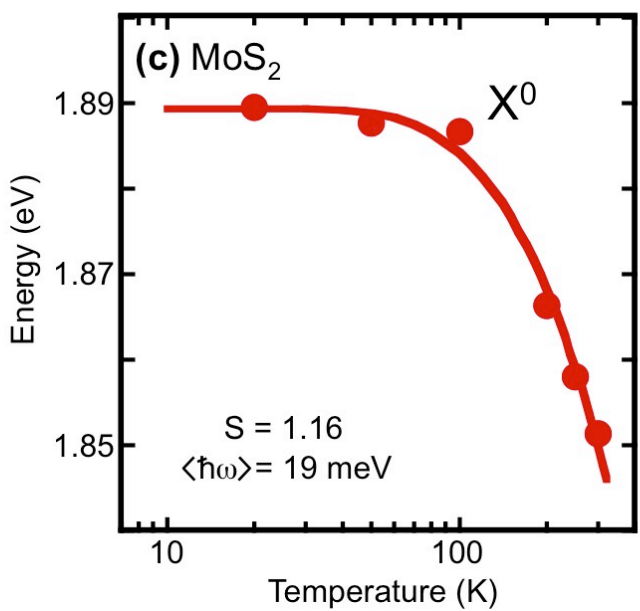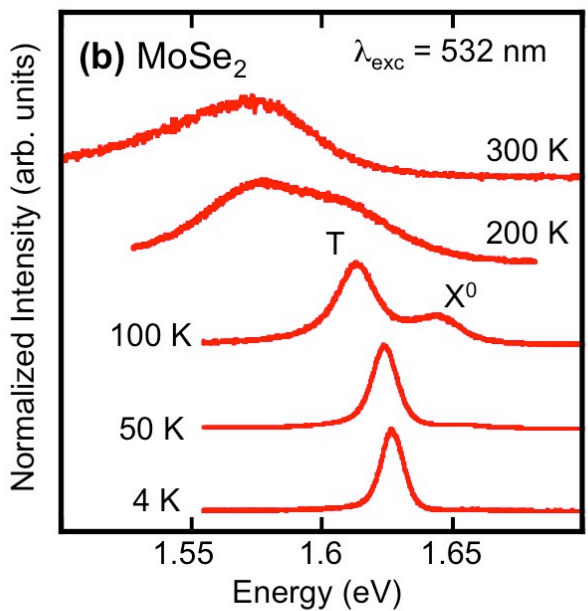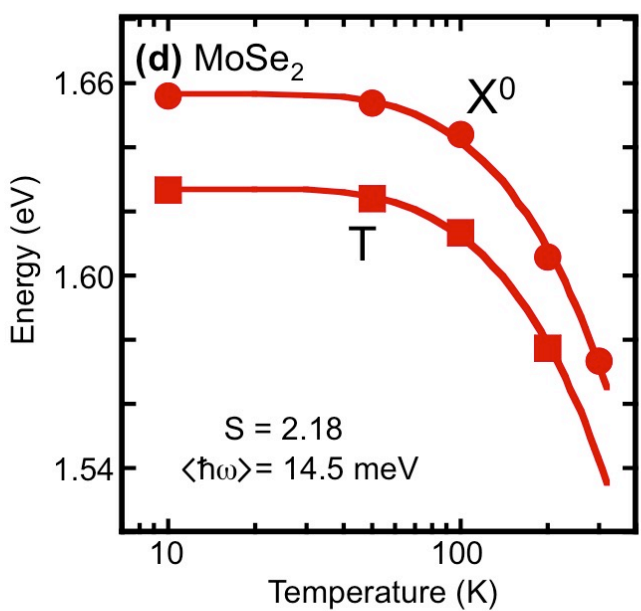

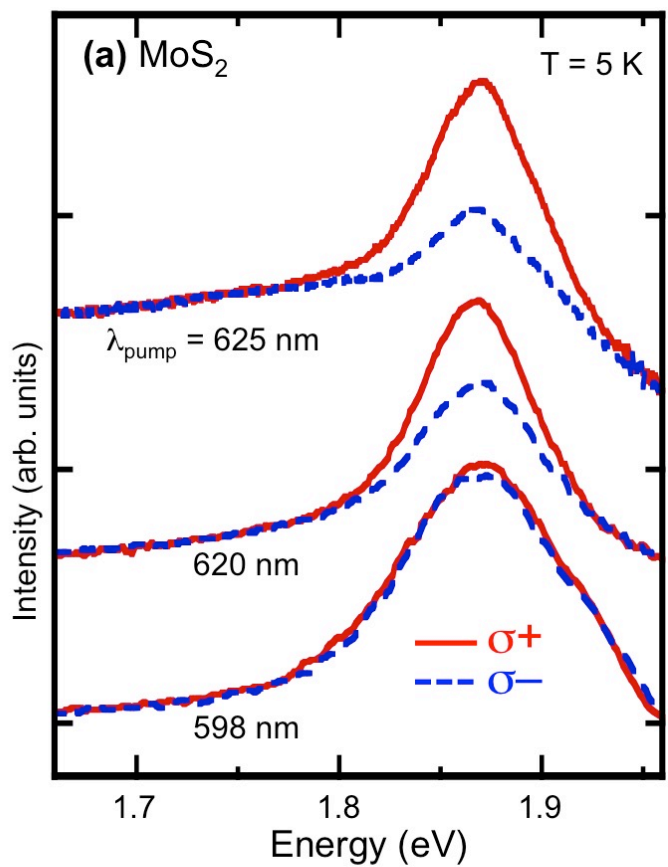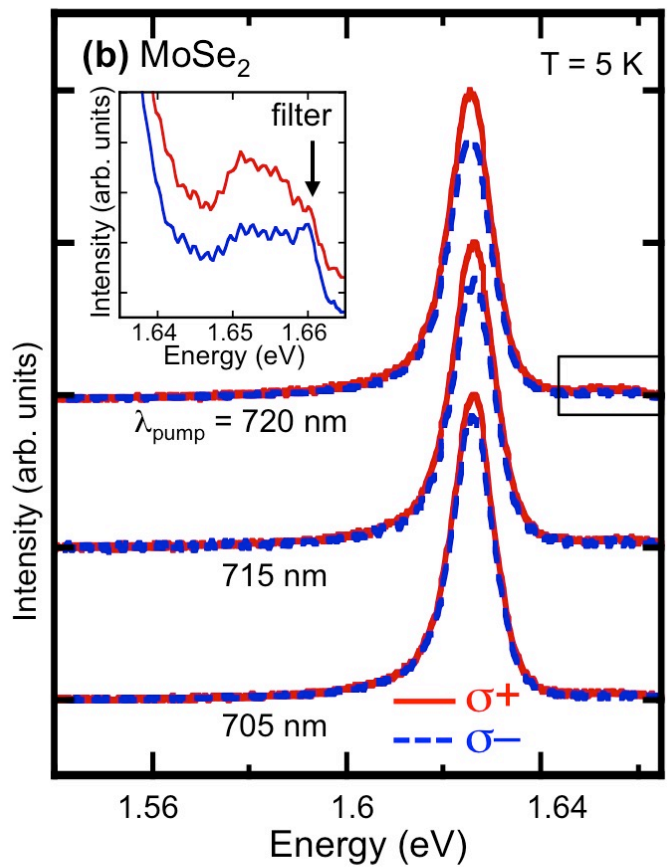

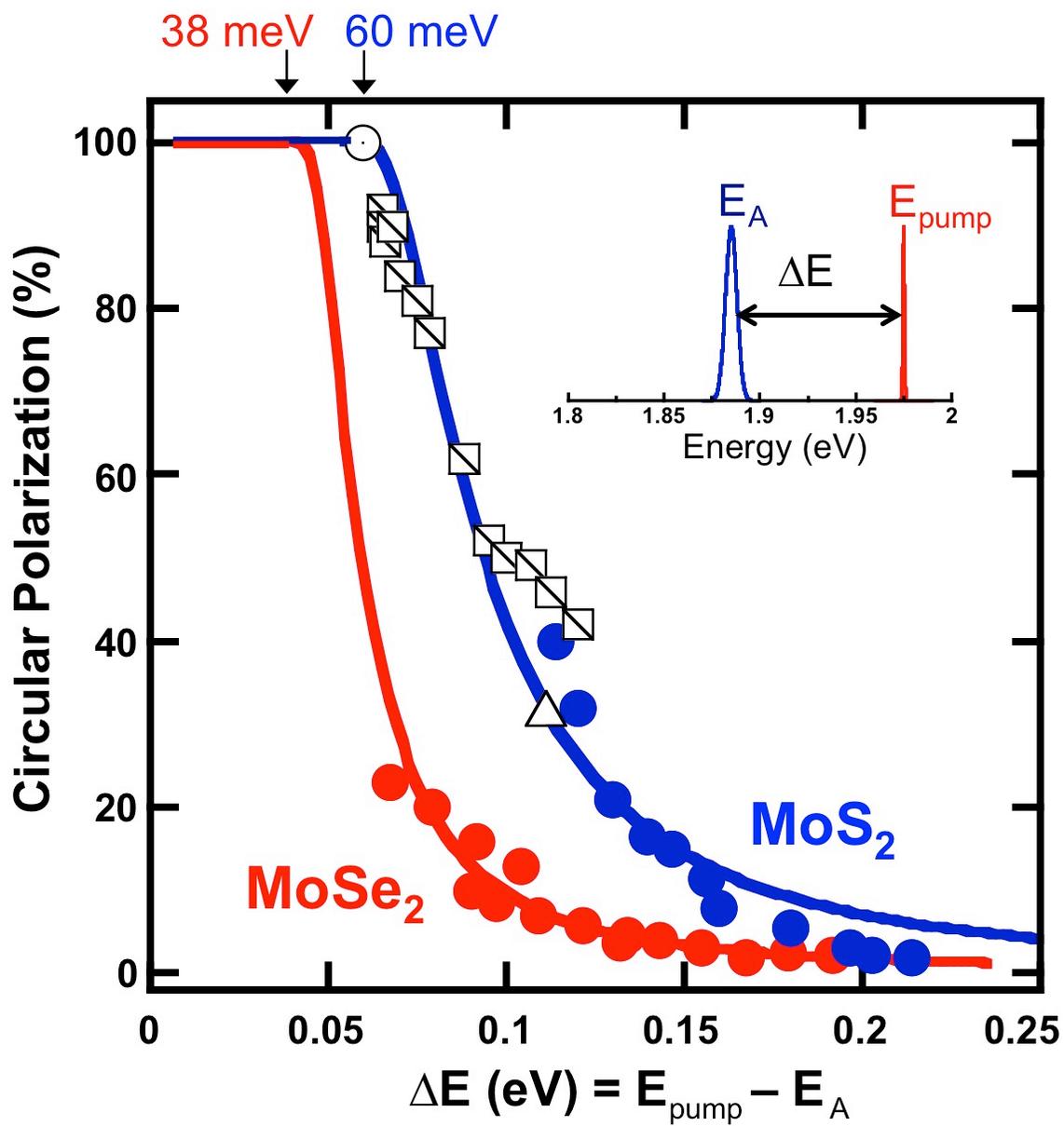